# Automated *Acanthamoeba polyphaga* detection and computation of *Salmonella typhimurium* concentration in spatio-temporal images

G D Tsibidis[1]♣ , N J Burroughs[2], W Gaze[3] and E M H Wellington[3]

[1]Institute of Electronic Structure and Laser, Foundation for Research and Technology, P.O.Box 1527, Vassilika Vouton, 71110 Heraklion, Greece

[2]Systems Biology Centre, University of Warwick, Coventry, CV4 7AL, UK

[3]Biological Sciences, University of Warwick, Coventry, CV4 7AL, UK

## Summary

Interactions between bacteria and protozoa is an increasing area of interest, however there are a few systems that allow extensive observation of the interactions. We examined a surface system consisting of non nutrient agar with a uniform bacterial lawn that extended over the agar surface, and a spatially localised central population of amoebae. The amoebae feed on bacteria and migrate over the plate. Automated image analysis techniques were employed to locate and count amoebae, cysts and bacteria coverage in a series of spatial images. Most algorithms were based on intensity thresholding, or a modification of this idea with probabilistic models. Our strategy was two tiered, we performed an automated analysis for object classification and bacteria counting followed by user intervention/reclassification using custom written Graphical User Interfaces.

---

♣ Corresponding author: tsibidis@iesl.forth.gr

**Introduction**

The importance of interactions between bacteria and protozoa has gained significance since the discovery that *Legionella pneumophila* could replicate within *Acanthamoeba polyphaga* (Stevens & O'Dell, 1973, Edelstein & Meyer, 1984). *Legionella pneumophila* is an organism which is strongly linked to the development of the Legionnaires' disease in humans (Edelstein & Meyer, 1984). The recognition of a connection between *Acanthamoeba polyphaga* and *Legionella pneumophila* (Stevens & O'Dell, 1973) has attracted much attention and possible association to other infections has also been investigated (Huws *et al.*, 2008).

*Acanthamoeba polyphaga* is a genus of amoebae, one of the most common protozoa, usually 15 to 35μm in length with a shape that varies from oval to triangular when moving. Due to its direct association to the development of various infections, *Acanthamoeba polyphaga* can function as a model organism used to analyse interactions between bacteria and protozoa. Elucidation of the underlying mechanisms entails the investigation of the amoeba behaviour in presence of bacteria. In principle, bacteria are ingested and digested by protozoa (Ekelund *et al.*, 2002) but some escape protozoan ingestion (Kinner *et al.*, 1998). Moreover, although some types of bacteria are ingested, they manage to escape and they even multiply within the protozoa. A typical example is the behaviour of the *Legionella pneumophila* which avoid digestion and multiply inside the organism (Molmeret *et al.*, 2005).

A powerful approach to the elucidation of the protozoan-bacteria interactions is to analyse the movement of protozoa in a bacterial lawn. Principally, the movement and behaviour is pertinent to the environment through two main processes, encystment under poor resources and bacteria dependent motility. A statistical analysis of the movement allows determination of the factors that characterise the underlying processes but usually a large number of organisms are necessary to infer useful information and draw a convincing picture. Video sequences that contain such organisms should be analysed to probe the protozoa behaviour (Gaze *et al.*, 2003), however, the large number of the organisms complicates the process towards a quantitative investigation. A large number of sophisticated methods based on neural networks (Casasent & Smokelin, 1994), wavelet transforms (Casasent *et al.*, 1992) and genetic algorithms (Kim & Jung, 2004), have also been widely employed and those state of the art techniques are designed to detect and analyse more complex images. The implementation of algorithms which are based on the above techniques are usually computationally more demanding. Moreover, subdivision of the original images into its constituent parts or objects requires the application of filters (Gerlich *et al.*, 2003), the employment of confinement trees and connected operators (Soille, 1999) or edge-based segmentation (Tvarusko *et al.*, 1999, Tvarusko *et al.*, 1998). By contrast, image analysis techniques based on image thresholding offer an adequate object detection method in pictures or video sequences characterised by a small number of objects with well defined boundaries. Several computer-assisted systems have been developed to track objects of biological importance in video sequences based on intensity thresholding (Sbalzarini & Koumoutsakos, 2005, Rabut & Ellenberg, 2004, Degerman *et al.*, 2009, Tsibidis & Tavernarakis, 2007). With respect to the detection of protozoans in lawns of bacteria, computational tools are required to be capable to facilitate object extraction, provide efficient classification (i.e. cysts or amoebae), determine quantitative details about the organism motility, count the bacterial concentration and offer a correlation of the bacterial distribution to the presence of organisms in the neighbourhood. Although most quantitative tools manage to extract organisms adequately and they are capable to identify protozoa in a series of images, they do not provide a satisfactory answer to how bacteria concentrations could be counted. Moreover, existing algorithms fail to characterise organism motility in two, temporally separated, images by a small time difference if the object boundaries are nearly immobile. Furthermore, specialised image processing methods are not generally applicable to the bacterial counting due to the difficulty to detect bacteria. Most commonly, the intensity of pixels in an image occupied by bacteria is very close to that associated with background which complicates the bacteria identification. As a result, bacteria enumeration would be a rather laborious process that requires a highly skilled expert. Previous works considered fluorescently tagged bacteria, a method that simplified the identification (Drozdov *et al.*, 2006, Pernthaler *et al.*, 2003, Singleton *et al.*, 2001, Soll *et al.*, 1988). Quantitative tools that address the above issues entail the modification of the existing algorithms with additional routines. The improved algorithms need firstly to manage the detection of the organisms and secondly allow an accurate classification of the two types of organisms in an efficient and robust way.

In the present paper, to overcome existing problems, an automated system has been developed aimed to assist the extraction of quantitative information of *Acanthamoeba polyphaga* and cysts in a lawn of *Salmonella Typhimurium* bacteria. Image processing techniques are employed to identify objects in a series of images based on thresholding. An automated classification of amoeba and cysts is firstly performed by considering a circularity criterion (i.e. cysts are almost circular). A manual correction of the object distinction is subsequently performed by means of a Graphical User Interface (GUI) which aims to integrate image processing and statistical classification technologies. A synergy of image processing techniques and probabilistic models for noise are subsequently used to assist in the bacterial counting. The proposed method is used as an alternative to the employment of fluorescent bacteria as a technique to facilitate bacterial detection. A similar approach is ensued towards the characterisation the organism motility. The algorithms developed in this paper intend to be an essential component of a completely automated system that will allow an adequate quantification of protozoa-bacteria interactions.

**Materials and methods**

*Strains*

The virulent strain of *Salmonella typhimurium* was used in co-culture experiments, and was grown in Luria Broth (LB). *Acanthamoeba polyphaga* CCAP 1501/18 was used as a model organism since *L.pneumophila* has been shown to grow in this strain. *A. polyphaga* was grown axenically in proteose peptone glucose medium (PPG).

*Co-culture experiments*

Bacterial cultures were washed 3 times in Page's Amoeba Saline (PAS) and diluted to approximately $10^8$ cells/ml (high bacteria concentration) and $10^7$ cells/ml (medium bacteria concentration). 100ml were spread onto non-nutrient agar (NNA) plates. Different approaches were tested to ensure uniform distribution of bacteria over the plates. It was found that by diluting the bacteria aliquot with 500ml of distilled water before spreading and allowing the liquid to dry on the plate a uniform distribution could be achieved. One week old cultures of *A.polyphaga* grown at $25^0$C were washed 3 times in PAS, and 25ml of approximately $6\times10^4$ cells/ml (1500 cells) was spotted in the centre of agar plates. Drop diameter and centre were recorded for each plate. Co-cultures were incubated at $35^0$C.

*Imaging*

Destructive sampling was carried out daily, with plates examined using a Zeiss Axioskop 2 microscope and 10X objective. Individual plates were used for measurements on days 0, 1, 2, 3, 5 and 7. Later time points were not required in bacteria free systems as they rapidly encyst. At each time point a single plate was used for each bacterial density (high, medium, none). A rectangular lattice of images was generated at a periodic spacing of 1mm covering part of the fourth quadrant measured from the centre of the amoeba solution drop (centre corresponds to lower right corner of first image). Lattice dimensions were adjusted depending on amoeba migration. Images were 1022 pixels x 1024 pixels (8bit), captured using a Hamamatsu black and white digital camera at magnification x10 and Improvision Openlab software. Each image covered approximately 0.46mm$^2$. No coverslip was used to preserve spatial integrity of the bacteria and amoebae populations.

*Amoeba-cyst location*

The refractive halo is used to locate amoeba/cysts by intensity thresholding; In principle, a threshold value equal to 0.85 (by converting gray scale image intensities in the range [0,1]) was chosen based on the fact that the great majority of the objects could be extracted adequately. The process yielded quantitative information associated to the centre of mass location, perimeter, size, area, and shape statistics. Because the intensity profile of parts of a small number of objects was very close to the threshold value, a manual perimeter correction was required for these objects (Figure 1a). Similarly, a small number of objects were not detectable by their halo and therefore they were added manually by means of a Graphical User interface designed in Matlab (Burroughs *et al.*, 2003).

*Amoeba-cyst classification*

Cysts have a nearly spherical shape and the projection of the organisms on the plane of focus yields an almost circular shape. By contrast, amoebae are usually characterised by an oval shape. Therefore, the *'circularity index'* (*CI*) of the object perimeter can be used as a criterion for the distinction between amoebae and cysts (Soll et al., 1988). It is defined by

$$CI = \frac{\delta r}{r} \qquad (1)$$

where *r* is the average radius of the object and *δr* is the standard deviation of the radius values. Therefore cysts should be associated with a small *CI*. In principle, a circularity index value is determined by the user in a way that allows an efficient amoebae/cyst distinction. It turned out that a value equal to 0.035 offered efficient object distinction, however, further modification was pursued based on clustering in the *radius* vs. *CI* scatter plot (Figure 2a). The employment of the latter technique aimed to improve results from automated classification in some cases.

*Mobile-Immobile analysis*

The distinction between mobile and immobile organisms was facilitated by the employment of a delayed image comparison. More specifically, for two observations of the same region ($\Delta t$=4secs apart), we define

$$I_d = I(t + \Delta t) - I(t)$$
$$<I> = \frac{1}{2}\left[I(t + \Delta t) + I(t)\right] \qquad (2)$$

where *I(t)* and *I(t+Δt)* are the intensities of images at time *t* and *t+Δt*, respectively. Due to fluctuations in photon counts, noise can be modelled as a Poisson variate, *i.e.* the pixels in images I*(t)*, *I(t+Δt)* should be Poisson distributed with mean *<I>* under the noise model. This suggests that the variable $f = \frac{I_d}{\sqrt{<I>}}$ is approximately Gaussian with zero mean, standard deviation $\sigma$ and it follows a distribution of the form

$$\frac{A}{\sqrt{2\pi}\sigma}\exp\left(-\frac{f^2}{2\sigma^2}\right) \qquad (3)$$

where *A* is a rescaling parameter. Thus, an intensity histogram of *f* for two consecutive images should be Gaussian when there is no movement. By contrast, if the organism moves significantly, some pixels inside the mask will be characterized by intensity changes greater than the values that can be described by the Poisson model. Parameters $\sigma$ and *A* of the distribution are estimated from the image intensity profile by a recursive algorithm that runs over the whole set of images. The above model yields an excellent fit to real data (Figure 3a) despite discretisation of image intensities requires a careful treatment when generating histograms. After subtracting out the noise from the observed distribution profile (i.e. *h(f)*), the probability that a pixel with observation *f* is unrelated to noise is computed from the expression

$$p(f) = \max\left(0, 1 - \frac{A}{h(f)\sqrt{2\pi}\sigma} \exp\left(-\frac{f^2}{2\sigma^2}\right)\right) \quad (4)$$

Figure 3b illustrates the probability $p(f)$ as a function of the intensity difference and it is remarkable that for pixels with large intensity changes the probability is very close to 1. The probability is restricted to be positive; moreover, noise subtraction is not exact due to random statistical fluctuations. Using a confidence level on $p(f)$ (i.e. $p(f)=20\%$), the number of 'mobile' pixels is counted inside the amoeba/cyst perimeter to remove fluctuations. We also define the ratio $R$,

$$R = \frac{\text{Maximum Pixel Value}}{\sigma} \quad (5)$$

where *Maximum Pixel Value* is the maximum intensity expressed by pixels inside the mask of the object and $R$ characterises the mobility of the organism. On average, mobile objects display greater intensity changes and contain a higher percentage of 'mobile' pixels.

*Bacterial coverage*

Bacteria analysis methodology was similar to the approach which was ensued to achieve mobile-immobile object distinction. The method was motivated by the observation that transmission variation through agar produced an approximate Gaussian agar intensity profile and darker pixels corresponded to bacteria. From the original intensity image $I$ (Figure 5a), an image $I_r$ was constructed where all objects were removed, *i.e.* all pixels inside the mask of the organism are ignored. Provided bacterial coverage was less than 30% of the plate, the intensity profile was dominated by the agar surface with an approximate Gaussian peak with mode near intensity levels 0.4-0.6, which varied from image to image because of refocusing and intensity adjustment during data collection. The agar intensity profile was estimated by modelling it as a Gaussian distribution and fitting parameters such as the mean $\mu$, the maximum $A$ and the standard deviation $\sigma$ of the distribution from a regression analysis on the derivative of the intensity profile over an appropriate range. In practice, the fit was weighted (biased) to the left of the mode of the distribution to optimise the fit in the region of agar-bacteria intensity separation. A probability of a pixel with intensity $I_r$ being bacteria is subsequently computed by subtracting the Gaussian agar intensity profile from the curve that corresponds to the observed data. Thus, results can be attained by computing the percentage of the area covered by bacteria, *BC*,

$$BC = \frac{B}{S-M} 100 \qquad (6)$$

where *B* stands for the number of pixels which are identified as bacteria (probability weighted provided *p(f)*>20%, threshold removes stochastic fluctuations), and *S-M* is the number of pixels which are not covered by objects (*S*: size of image in pixels, *M*: number of pixels occupied by objects).

**Results and Discussion**

Basic image analysis techniques based on intensity thresholding were firstly used to locate and count amoebae and cysts. Image processing algorithms were applied to all datasets aimed to identify all objects and determine a set of morphological details such as position, shape, perimeter and area. Small objects of no importance were removed from consideration by adapting the algorithm to ignore tiny, light regions (*i.e.* comprising fewer than 20 pixels). Based on object outlines, a quantification model was developed. Results revealed that a small number of organisms were not detected correctly by the algorithm or equivalently, the automated system resulted in a considerably overestimated number of objects. Time consuming manual classification for every individual object was avoided by running an automated classification procedure based on the circularity index criterion to distinguish amoebae from cysts.

Our strategy was two tiered, an initially automated analysis was performed for classification and counting, followed by user intervention/reclassification using a Graphical User Interface (GUI) designed in MATLAB (The Math Works, Natick, MA). The GUI (Fig.8) was introduced to overcome problems related to classification errors, incorrect perimeter estimation, missing objects due to small intensity contrast with respect to background and inadequate separation of multiple objects situated in a very small region. The interface allows the user to intervene and correct interactively classification and morphological errors. All manual changes are recorded and overall analysis is adapted to a modification which makes the GUI a dynamic interface. Moreover, the GUI contains viewing tools to obtain data, measurements, confidence statistics and images for individual objects that allow the acquisition of summary information for individual objects or whole categories of objects. The analysis tools are employed to facilitate a fast quantitative and qualitative investigation for objects by enabling biologists to view plots of results (Figs1-4) (Burroughs *et al.*, 2003).

A secondary classification was ensued with the employment of clustering algorithms applied to data in the scatter plot *δr/r* vs. *r* (Fig.2a); due to small values of *δr/r*, cysts were clustered in the lower section of the graph and objects were thereby classified based on approximate clustering. The proposed classification algorithms that are components of the GUI, facilitated the creation of clustering patterns that allowed a faster classification and decision making process. Moreover, the viewing facilities of the interface provided a rapid scan of a library of objects belonging to a particular category, allowed a comparison and offered the capability of organism reclassification. Fig.2a illustrates a handful of objects that can be identified as amoebae despite they reside inside the cluster and have a nearly circular shape. In this case, a manual classification is required because amoebae are immobile and highly spherical when they are in the division phase. As a result, amoebae and cysts are indistinguishable and they are identifiable only through visual inspection of the image. The advantage of the method, however, is pertinent to the minimisation of the need for a manual object classification of the entire data set. It is important to emphasise that the above object classification criteria have been applied to objects with a well defined boundary. By contrast, amoebae and cysts with a part outside the image are either excluded from consideration or indentified manually by observation (Fig 1a). Similarly, analysis was not performed on a small number of unidentified fragments (*eg*. Agar crystals) that were removed from consideration.

To distinguish mobile from immobile objects and facilitate cytoplasmic movement, delayed images $I_d = I(t + \Delta t) - I(t)$ were generated (Figure 1b) and $I_d$ values were fitted with a Gaussian distribution (Figure 3a). A probabilistic computation of the percentage mobile pixels in every object (Figure 3b) was subsequently used to identify mobile objects, specifically through clustering in the scatter plot *R* vs. *percentage of mobile pixels* (Figure 2b). Clustering in conjunction with examination of objects in the bordering areas was performed by the employment of the GUI which allowed a rapid and efficient determination and characterization of all moving organisms. Separate mobile-immobile classification regions were explored for amoebae and cysts since cytoplasmic movement was large in amoebae, in contrast to movement inside cysts. Therefore mobile-immobile classification for cysts was less reliable, however, the ecosystem dynamics is not influenced due to that, because the role of cysts is of minor importance.

A powerful approach to the elucidation of the protozoan-bacteria interactions is associated to a thorough investigation of the organism number dependence on bacteria distribution. A consistent elucidation entails bacteria counting that are derived by analysing sets of images based on experiments with organisms in lawns of bacteria of various concentrations (low, medium, high). Without loss of generality, for the inoculation procedure (Figure 4), a drop of amoebae on a lawn of bacteria (medium concentration) was used and the distribution of amoebae and cysts was investigated during several days. Figure 4a and Figure 4b illustrate the position of amoebae-cysts on days 2 and 3, respectively, in the first quadrant. On day 2, there exist 56 amoebae (50 mobile and 6 immobile) and 50 cysts (22 mobile and 28 immobile) and most of them are situated inside the drop radius (i.e. 11mm). By contrast, on day 3, the number of amoebae has increased to 160 (156 mobile and 4 immobile) while the number of cysts has essentially remained unchanged (49). There are also 11 organisms that cannot be classified to any of the above categories even by the expert's eye. The unclassifiable objects are usually amoebae in the process of division or single organisms made up of two or three amoebae ('multiple objects'). 'Multiple objects' constitute a set in the 'All others' category in Table 1. Other types that are classified in the 'all others' category include bacterial colonies, carcasses, remnants (i.e. unidentifiable objects), other unclassifiable organisms. It is evident that most of the mobile objects have moved out of the drop radius on day 3 (i.e. 15mm) which is an indication that in presence of bacteria, very few amoebae encysted; By contrast, in bacteria-free plates (data not shown), the majority of amoebae encysted within 3-5 days with little migration out of the original drop. Moreover, amoebae demonstrate a wide variation in size (Figure 6A), which does not correlate with distance of migration.

Spatial bacterial coverage was also calculated by means of a probabilistic approach described in Materials and Methods. It is assumed that $I$ is a matrix that represents the original image (Figure 5a) before an intermediate modification is introduced by masking out the pixels covered by all objects in the image. Special attention is required for dark pixels inside the organism due to the fact that the associated intensity is close to the value that corresponds to the bacterial intensity. Therefore, to avoid overestimation of bacteria counting, the mask of the objects is removed. A new matrix $I_r$ is derived and $I_{min}$ and $I_{max}$ are the smallest and largest values in the intensity histogram $I_r$ and $r \in [I_{min}, I_{max}]$. It is assumed that noise in the image follows a Gaussian distribution G (Figure 6b) and a quantity $p_r$, equal to $\frac{r-G}{r}$, is assigned to every intensity value $r$. The physical interpretation of $p_r$ is that it represents the probability that a pixel in the image with intensity $r$ corresponds to a bacterium, for pixels darker than the agar. There is a lower bound $r_0$ which is a cut-off value over which image pixels fail to contribute to the bacteria density. The number of bacteria pixels in the image is provided by

$$\sum_{r=I_{min}}^{I_{max}} p_r(r-G) \qquad (7)$$

The peak of the intensity values *(red* line) in every image was fitted with a Gaussian distribution *(blue* line) that yields the bacterial curve *(green* dotted line). The Gaussian distribution parameters $\mu$ (mean) and $\sigma$ (standard deviation) were calculated for all images in the datasets through an automated procedure and subsequently, bacteria coverage was calculated along with the spatial bacterial distributions. All dark pixels that correspond to bacteria have been painted in red (Figure 5b). Occasionally, when the automated fit with a Gaussian fails, a fit on a different range of intensities (i.e. not in the whole range) is performed and if the procedure still leads to poor results, either a manual threshold is selected or the image is simply removed from consideration. Then, pixels with intensity less than the threshold were regarded to be bacteria. In principle, the efficiency of our approach was very high for bacterial lawn with medium density and a very small number of images required special treatment. The latter often occurred for images characterised by uneven illumination; regions in the centre of the image systematically exhibited a higher degree of brightness, by approximately 10% relative to the edges caused by the microscope system. A normalization procedure was followed to remove the discrepancy; more specifically, every image with intensity $I$ was replaced by a normalised image with new intensity $I/D$, where $D$ represents the average of all images in every set. If this manipulation, though, failed to provide a satisfactory bacteria counting either the entire images or problematic regions of pictures were removed from the analysis. Findings attained by the above method demonstrated that the percentage of images that turned to be problematic and failed to lead to acceptable results was extremely small (about 2%).

A spatio-temporal analysis was also conducted for bacteria coverage on several days (plots are illustrated for days 2 and 3, respectively) and results demonstrated decrease of the bacterial density due to amoebal grazing (Figure 7). This behaviour is indicative of a correlation of the amoeba density and reduced bacterial coverage. Each image in the lattice was represented by a coloured box where the colour corresponded to the bacterial coverage. The uniform colouring of each box does not suggest that the bacterial coverage is constant in the whole image but the particular representation of the spatial bacterial distribution is attributed to simplicity reasons. The white lines in Figure 7a and Figure 7b represent the original drop radius.

In the evaluation procedure, the main objective was to assess a set of requirements including rapid and accurate object recognition and reliability against the decision of experts. The accuracy of the tool comprising the automated object/mobility classification procedure and GUI was tested. It is evident that for any automated system to be of use and merit, apart from its ability to be fast in the extraction and classification of objects, it must be trustworthy. A convincing approach to estimate the reliability of the technique necessitates an examination in various imaging conditions and a quantitative measurement of its performance in comparison with a human expert decision. Object extraction and classification was performed in all image datasets containing organisms in bacterial lawns of various densities (low, medium, high). Depending on the number of objects in every image, computation time to extract all objects varied. Nevertheless, the automated procedure to obtain morphological, mobility and type details did not require longer than ten seconds for every image (fifteen organisms was the maximum number of objects in every image). To illustrate the validity of the proposed method, Table 1 illustrates the number of organisms in a lawn of bacteria of medium density during a period of five days. Furthermore, reliability of results was evaluated against the decision of experts and results due to visual reclassification and manual morphological intervention (i.e. perimeter redrawing) are also presented in the same table. Table 1 shows excellent correspondence between the two sets of results. Only a few objects required manual reclassification or perimeter redrawing by the expert (i.e. morphological operation entailed less than 0.5secs/object). In principle, conflicting results were only noted when organisms lacked features that enabled a consistent and objective computer-aided or expert judgement. The efficiency of the machine performance suggests that the combination of the automated system and the clustering algorithms constitutes a trustworthy technique that provides precise measurements and minimises manual intervention. Unlike other intensity thresholding based methodologies (Sbalzarini & Koumoutsakos, 2005, Rabut & Ellenberg, 2004, Degerman et al., 2009, Tsibidis & Tavernarakis, 2007), our technique utilises a very robust probabilistic method of assessing mobility of objects. The key impact of the approach is significantly useful in cases where object outline is not substantially mobile while interior parts of the organism move. Furthermore, viewing facilities of the GUI offered assistance to acquire a fast overview of the results while morphological correction functions facilitated manual intervention. The proposed system is characterised by an increased computational capability for precise extraction and classification of amoebae/cysts in lawns of bacteria of low and medium density. By contrast, significant improvements are required for the analysis of sets of images that contain organisms in lawns of bacteria with high density. Evolution of bacteria colonies produce regions of high intensity which are mistakenly regarded as parts of organisms.

|  |  |  | **Data** | | | | |
|---|---|---|---|---|---|---|---|
|  |  |  | Day 1 | Day 2 | Day 3 | Day 5 | Day 7 |
| **Type of objects** | Mobile Ameobae | Automated | 50 | 156 | 180 | 152 | 160 |
|  |  | Reclassified | 1 | 8 | 12 | 10 | 9 |
|  | Immobile Amoebae | Automated | 6 | 4 | 22 | 15 | 22 |
|  |  | Reclassified | 0 | 0 | 2 | 4 | 6 |
|  | Mobile Cysts | Automated | 22 | 29 | 55 | 47 | 38 |
|  |  | Reclassified | 1 | 1 | 5 | 3 | 4 |
|  | Immobile Cysts | Automated | 22 | 20 | 42 | 35 | 39 |
|  |  | Reclassified | 3 | 0 | 4 | 4 | 8 |
|  | All others | Automated | 47 | 60 | 86 | 51 | 61 |
|  |  | Reclassified | 3 | 13 | 15 | 5 | 10 |
|  | Redrawn Objects | - | 13 | 30 | 32 | 24 | 28 |

**Table 1**: Objects obtained through the synergy of the automated procedure and the clustering algorithms on various days. Comparison of results from computer assisted methods and manual intervention is also presented by reference to the number of organisms that needed manual reclassification

In regard to the enumeration of bacteria, the applicability of the probabilistic algorithm was tested against visual estimation in bacterial lawns of various densities,. Unlike previous reports and automated systems (Jansen *et al.*, 1999, Grivet *et al.*, 2001, Singleton et al., 2001), our method avoids the need for training of the system and it offers a precise computation of the bacterial coverage. For low and medium bacterial coverage about 95% of the images were analysed efficiently and realistic computations of the bacterial coverage was performed. By contrast, bacterial analysis in areas characterised by a higher concentration is problematic and either a tedious manual or an alternative methodology is required. Therefore, significant modifications and multiple selective decision rules need to be incorporated into the system to improve the efficiency of the computer-aided bacterial computation. Nevertheless, the proposed method for bacterial spatial characterisation appears to be highly suitable for a quantitative and qualitative investigation of systems characterised by a non high bacterial density. Furthermore, the technique represents an early initiative for a systematic approach of modelling the ecology of *Salmonella typhimurium* in various environments. Although analysis demonstrates that stochastic behaviour at the level of individual has significant impact on spatial heterogeneity at the population level, some clear dynamic trends are apparent.

At present, the mechanism involved in the bacterial decrease due to grazing is not well understood and requires further investigation. Protozoan population regulation is the major factor governing encystment and interestingly, the rate of encystment is controlled by a bacterial density dependent mechanism. In principle, determination of the protozoan-bacteria dynamics can be modelled with reaction diffusion equations (Ekelund et al., 2002) which confirms the significance of the knowledge of the bacterial identification. The results obtained through the proposed probabilistic approach demonstrate that the system can sample interactions of protozoans with bacteria with high accuracy and it can constitute a major component of any improved analysis tool.

**Conclusions**

We have developed a widely applicable technique for accurate analysis and elucidation of amoebae – bacteria interactions. In the first application, amoebae and cysts were localised on lawns of a varied bacterial density and morphological quantities were obtained. The method proved to be particularly useful because it allowed an automated analysis with a remarkably high efficiency and reliability. A second technique was additionally presented that aimed to compute the bacterial coverage in lattice experiments. The method is also characterised by high efficiency and it demonstrates the decrease in the bacterial density due to amoebal grazing. This finding indicated a correlation between amoeba density and reduced bacterial coverage. The evaluation procedure demonstrated that the efficiency of the system was remarkably high and it constituted a competent method to obtain information that can elucidate protozoa-bacteria interactions, migration and growth events in an ecosystem at both the individual and population levels. Moreover, our methodology could be also applied to other biological systems that exhibit a similar behaviour.

**Figures**

**Figure 1.** (a) Image that contains amoeba and cysts. *Blue* lines represent the perimeter of each object, *red* dots indicate their centre of mass and light *blue* lines stands for perimeter in case manual correction was performed, (b) Intensity image of delayed image $I_d = I(t+\Delta t) - I(t)$.

**Figure 2.** (a) Clustering procedure to simplify amoeba-cyst classification. The region indicated by the *blue* polygonal line has been drawn manually to include objects that are classified as cysts. *MA, IA, MC and IC* stand for mobile amoebae, immobile amoebae, mobile cysts and immobile cysts, respectively, (b) Clustering procedure to simplify mobile-immobile amoebae classification. The manually drawn region denoted by the *blue* polygonal line indicates objects initially identified as immobile amoebae.

**Figure 3.** (a) Gaussian fitting of variable $I_d$, (b) Distribution of probability values that a pixel with temporal intensity change $I_d$ is unrelated to noise.

**Figure 4** (a) Distribution of amoebae and cysts on day 2 on medium density bacterial lawn (drop size is 11 mm), (b) Distribution of amoebae and cysts on day 3 on medium density bacterial lawn (drop size is 15 mm). The dotted *blue* line indicates the original drop size. 'MO' corresponds to objects that are unclassifiable by observer due to the fact that they contain multiple objects.

**Figure 5.** (a) Original image of bacteria (*dark* spots) in the presence of protozoa, (b) Image $I_r$ covered with bacteria (*red* spots).

**Figure 6.** (a) Amoeba radius size on day 3, (b) Fitting of the peak of the intensity values (*red* line) with a Gaussian distribution (*blue* line) to compute noise and derive the bacterial curve (*green* dotted line). Gaussian distribution parameters were calculated ($\mu$=0.3931 and $\sigma$=0.019).

**Figure 7.** Bacterial coverage for days 2 (a) and 3 (b), respectively. Colour shows percentage of image area covered by bacteria.

**Figure 8.** Graphical User Interface that allows manual perimeter correction and reclassification of objects. (Figures in parentheses represent numbers of reclassified objects).

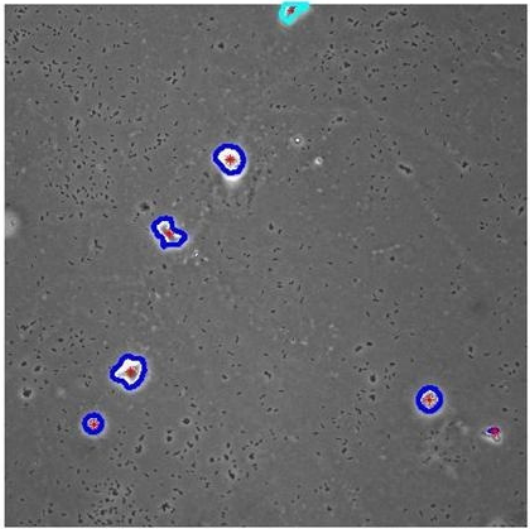 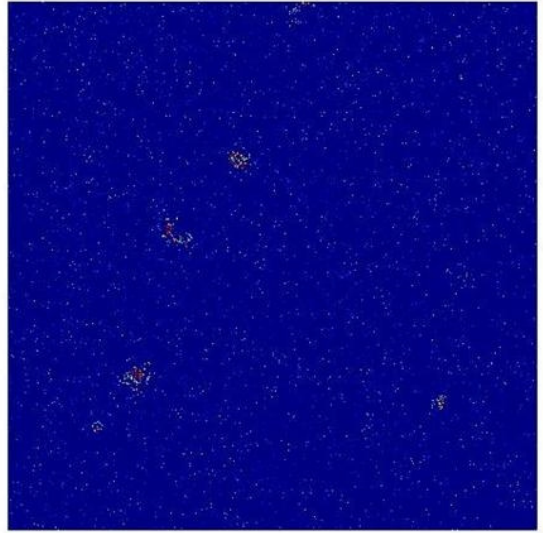

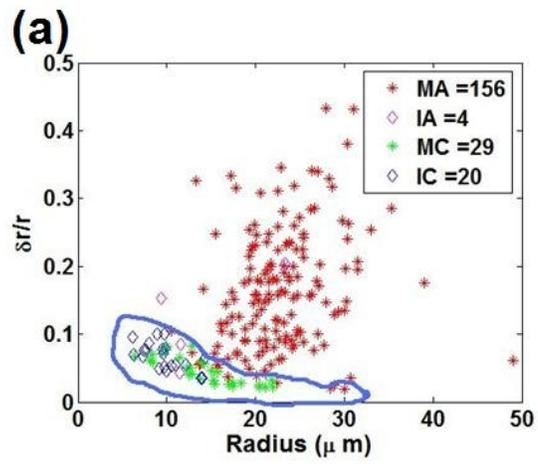 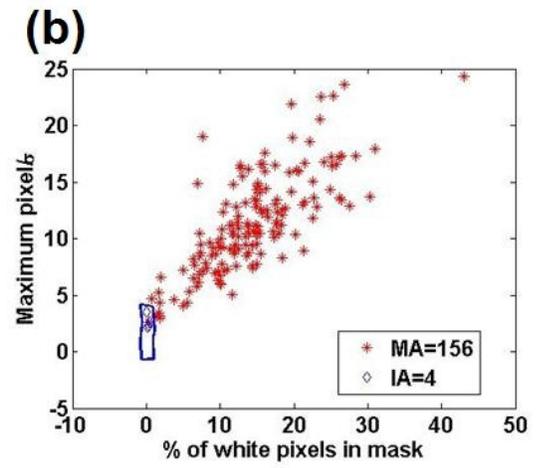

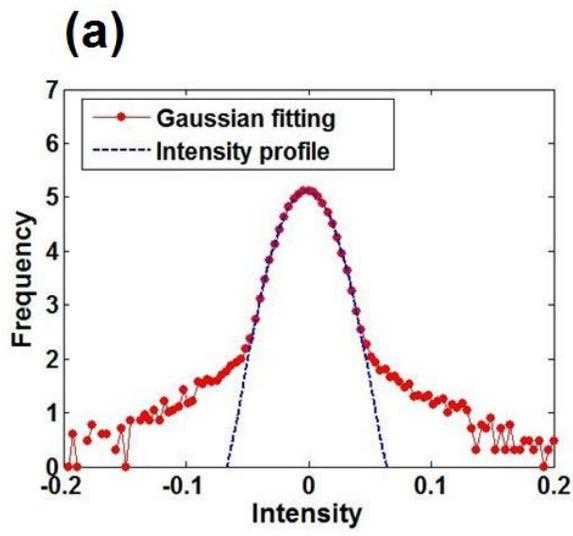 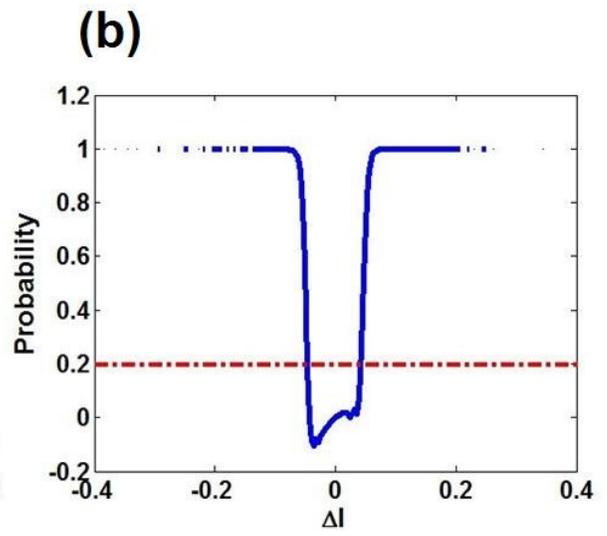

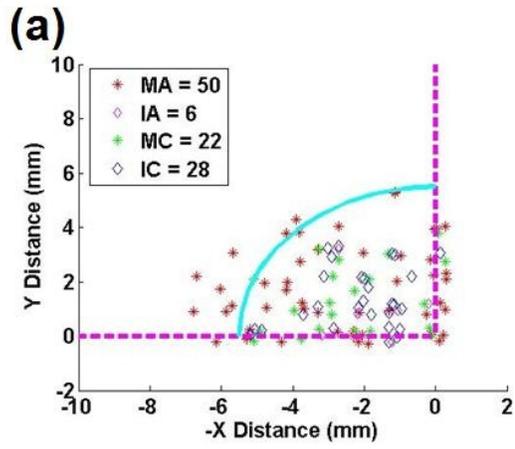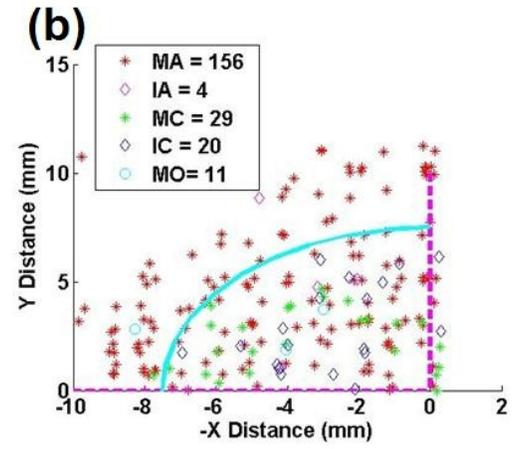

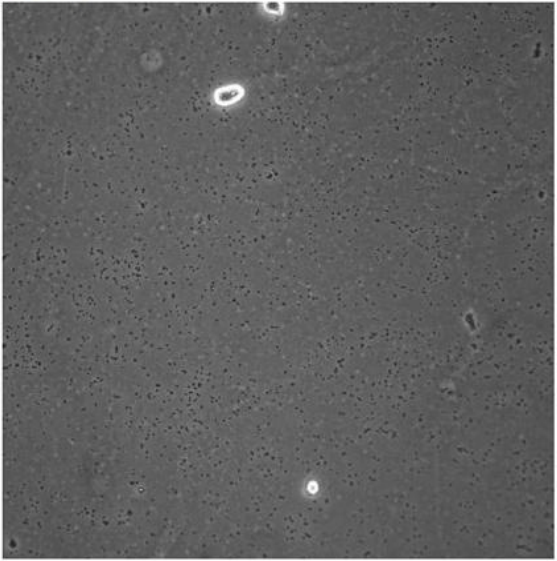 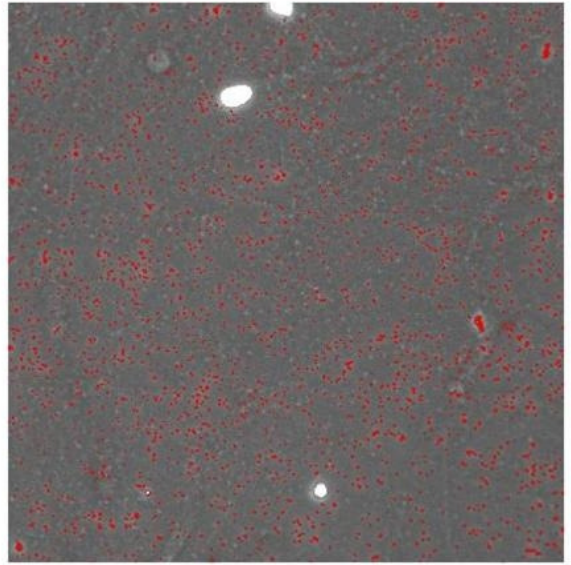

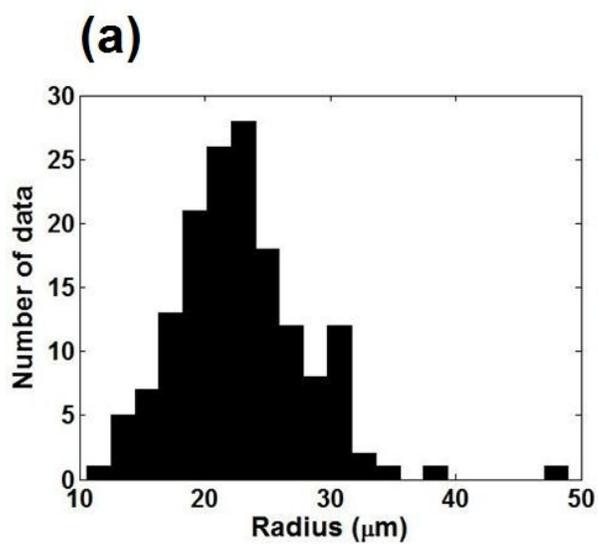 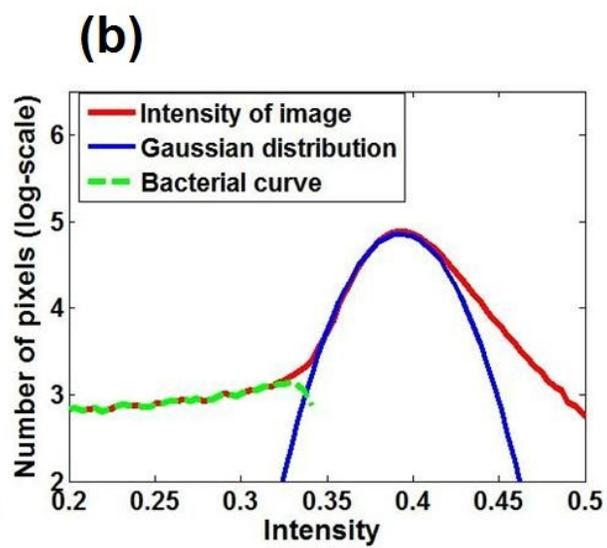

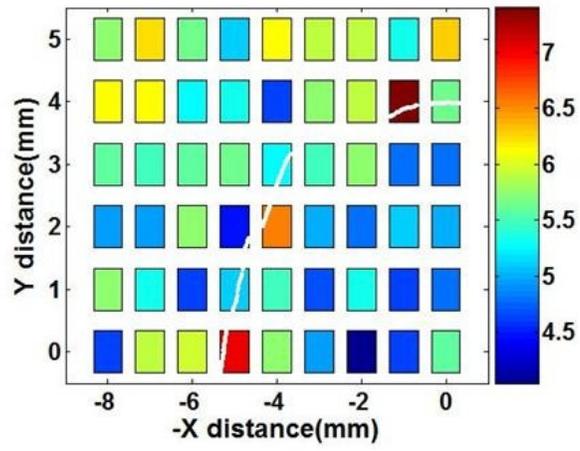 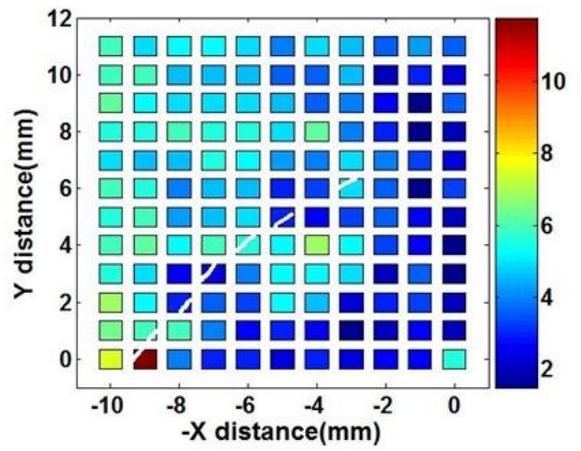

GUI screenshot showing analysis software panels for object classification (amoebae, cysts, etc.).

Panel 1 (top-left): medbac_2709
- Statistics for all objects | Prepare data
- Number of objects = 269 (22)
- Graphs
- X-Y coords of COM | 5mmu<r<15...
- # bins (default=100) | dr/r for all objects
- Classification: ○ auto ● reclass
- Threshold Cyst:<= 0.03500 Unknown:>= 0.40000
- D-L Class: Amoeba= ROI, cysts= ROI, rest objects= ROI
- CHANGED

Counts table:
| | # All | Live | Dead |
|---|---|---|---|
| # All = | 269 (22) | 227 (11) | 42 (11) |
| # amoebae = | 160 (8) | 156 (8) | 4 (0) |
| # cysts = | 49 (1) | 29 (1) | 20 (0) |
| # edge = | 15 (0) | 12 (0) | 3 (0) |
| # unknown = | 16 (4) | 9 (2) | 7 (4) |
| # remnants = | 10 (1) | 2 (0) | 8 (3) |
| # Colonies = | 0 | 0 | 0 |
| # Carcasses = | 13 (2) | #NOT obj = 2 (2) | |
| # multiple = | 4 (0) | # reshaped = | 30 |

- Images-graphs: all-live
- Cy-Am: all-live(del-rgb)
- Image = 6 (-2)< >(+2) (-6)<< >>(+6)

Panel 2 (middle):
- Choose image = 6 (between 2 - 265) Images
- Only the Image
- Number of objects = 3,2

| | Live | Dead |
|---|---|---|
| # All = | 3,2 | 1,0 | 1,0 |
| # amoebae = | 1,0 | 1,0 | 0,0 |
| # cysts = | 1,0 | 0,0 | 1,0 |
| # edge = | 0,0 | 0,0 | 0,0 |
| # unknown = | 0,0 | 0,0 | 0,0 |
| # remnants = | 0,0 | 0,0 | 0,0 |
| # colonies = | 0,0 | 0,0 | 0,0 |
| # Carcasses = | 0,0 | | |
| # multiple = | 1,0 | # reshaped = | 1,0 |

- Object label = 2
- radius = 30.383000 +/- 6.110000 mum
- Position = x-coordinate -1.89197 mm, y-coordinate -0.05130 mm
- Perimeter =
- Classification: 1(L-A) ,1(L-C) | Initial Classification: Live
- Manual reshape? No | Before reshape
- Object | Object in Image | X-Y position
- x -

Panel 3 (right):
- Correct shape | delete reshape
- New object | Remove all new objects
- Add new object | Object = *
- Correct shape of new object
- delete reshape of new object

- Reclassification of object
- Image = 6
- Object label = 2
- Type of object = amoeba
- Dead-alive = live
- Multiple Object: L-Am | De-Am | L-Cy | De-Cy
- Save info

- New Statistics for objects
- All Objects (Am+Cy)
- X-y coords of COM
- # bins (default=100)
- X-axis: radius | Y-axis: radius
- Display graphs | Draw ROI (dead-al...)
- Object